\documentclass[journal]{IEEEtran}

\usepackage{graphicx}

\begin{document}

\title{Exploring the efficacy of molecular fragments of different complexity in computational SAR modeling}
\author{Albrecht Zimmermann, Bj{\"o}rn Bringmann, Luc De Raedt
\thanks{Albrecht Zimmermann, Luc De Raedt, KU Leuven, firstname.lastname@cs.kuleuven.be}
\thanks{B. Bringmann, Deloitte \& Touche GmbH, bbringmann@deloitte.de}}

\maketitle
\newcommand{\ea}[1]{#1~\emph{et~al.}}
\newcommand{\algo}[1]{{\sc #1}}
\newcommand{\fp}{\emph{fp}}
\newcommand{\gfp}{\emph{gfp}}

\begin{abstract}

An important first step in computational SAR modeling is to transform the compounds into a representation that can be processed by predictive modeling techniques.
This is typically a feature vector where each feature indicates the presence or absence of a molecular fragment. 
While the traditional approach to SAR modeling employed size restricted fingerprints derived from path fragments, much research in recent years focussed
on mining more complex graph based fragments. Today, there seems to be a growing consensus in the data mining community that these more expressive fragments should be more useful. 

We question this consensus and show experimentally that fragments of low complexity, i.e. sequences, perform better than equally large sets of more complex ones, an effect we explain by pairwise correlation among fragments and the ability of a fragment set to encode compounds from different classes distinctly. The size restriction on these sets is based on ordering  the fragments by class-correlation scores. In addition,  we also evaluate the effects of using a significance value instead of a length restriction for path fragments and find a significant reduction in the number of features with little loss in performance.

\end{abstract}


\section{Introduction}

Structure-activity relationship (SAR) prediction is an important task in computational biochemistry.
The 	aim is to predict the effect of compounds based on their
 structural characteristics 
 --~the second-order representation comprising the topological arrangement of atoms and bonds of the molecule.

For algorithms to 
process molecular data and build models to predict their activity,  
molecules have to be simplified by transforming them into a different representation. To this end, molecules are abstracted as \emph{graphs} -- networks of atoms linked to each other. A common approach to SAR consists of constructing fragments from individual or pairs of  molecules, and subjecting those molecules to \emph{fingerprinting} to gain a final representation that is more easily accessible for prediction algorithms such as \emph{Support Vector Machines}. The \emph{graph mining community}, on the other hand \cite{Washio:2003:SAG:959242.959249}, approaches the construction of fragments slightly differently and while the differences are subtle, they can have significant effects.

In this paper, we build on earlier work \cite{baldi} that aimed at generalizing the existing fingerprinting approach and explore
how to derive the molecular fragments on which to base generalized fingerprints in a predictive setting.
Specifically, we compare fragments of different complexity  in terms of their usefulness.
Additionally, we compare the use of fragments selected according to their correlation with the target value to ones selected using a threshold on their length.

The paper is structured as follows: we first discuss the concept of fingerprints and its extension to \emph{generalized fingerprints}, as well as differences in the complexity of fragments on which to base them. Following this, we lay out our methodology for the experimental comparisons in terms of complexity and selection criterion, on which we report afterwards. Finally, we discuss the observed phenomena and draw conclusions.

\section{(Generalized) Fingerprints}
\label{fingerprints}
The usual approach in computational biology/chemistry when using the second-order representation for SAR
involves assessing the structural similarity of
 molecules. 
They are decomposed into sets of (potentially overlapping) fragments and the similarity of any two molecules evaluated comparing their respective fragments using \emph{kernel functions} \cite{Haussler99convolutionkernels}.


A variety of different fragments has been used in the literature, from paths/walks \cite{DBLP:conf/colt/GartnerFW03,baldi,DBLP:conf/icdm/BorgwardtK05}, via fragments with branches (trees) \cite{DBLP:conf/nips/ShervashidzeB09}, to those with cycles (graphs) \cite{DBLP:conf/icml/MenchettiCF05,DBLP:journals/kais/WaleWK08,DBLP:conf/icml/CostaG10,DBLP:journals/ml/SchietgatCRR11}. Often, a new kernel function is proposed as well.
These approaches share two characteristics: 1) they start from vertices (atoms) of individual or pairs of molecules, enumerating the paths starting from this vertex, or the neighborhood graphs surrounding it. 2) the fragments are size restricted, length restricted for paths (such as $0 \leq l \leq 8$ or $3\leq l \leq 10$), or diameter restricted for graphs. 

The resulting fragments are often used to map molecules to bit-vectors of a given size $k$ (such as 512 or 1024), in a process called \emph{fingerprinting}, involving the generation of $b$ random integers that are mapped 
using a modulo $k$ reduction. 
While values such as $k$, $l$, and $b$ 
are based on empirical knowledge of biochemical practitioners, 
it has been shown that, e.g., different length restrictions can have a profound effect on the usefulness of derived fragments \cite{doi:10.1021/ci700052x}.
%

As an alternative to hashing, \ea{Swamidass} \cite{baldi} proposed \emph{generalized fingerprints} (\gfp) in which the \emph{explicit} size restriction on bit-strings is lifted. Thus, potential loss of information is avoided since \emph{each} fragment is represented. Also, it becomes possible to use information that goes beyond presence, e.g. how often a fragment occurs in the data, which the authors exploited in proposing a kernel. However, hashing fragments to the same bit can
weed out redundancy and such a representation 
potentially avoids the curse of dimensionality, and reduces memory requirements for the modeling step.
The retention of information seems to outweigh the benefits of hashing, since \ea{Swamidass} showed that \gfp{} outperform \fp{}, especially for smaller \fp{}s.

Their approach used path fragments, and in their conclusion
 they suggested the use of shallow trees as fragments from which to construct \gfp. 
This coincides with trends in the data mining community where \emph{graph mining} is touted as the tool of choice to derive fragments for SAR prediction.
In contrast to this there have been claims that 
 simpler features
may well suffice 
\cite{helma,DBLP:journals/jcisd/HelmaCKR04}. This assumption has been supported by recent work \cite{DBLP:journals/kais/WaleWK08} that evaluated the efficacy of structures of different complexity against one another and found little, if any, advantage in using more complex structures such as graphs. It has to be remarked, however, that the latter work still constructs \emph{size restricted} fragments from individual molecules.

In contrast to this, we have found in the past that sequential fragments are \emph{more} useful than more complex ones \cite{simpler-patterns}. Yet we construct fragments differently: they are not size restricted but chosen based on how well they correlate with the target variable, measured by $\chi^2$, a correlation that is evaluated on the entire data. The size restriction on fingerprints can either be enforced explicitly by taking the $k$ best-correlating fragments, or implicitly by requiring a minimum correlation score. 

 
%
%
We reproduce our experiments 
 on new data and perform additional analysis to answer the following questions:
\begin{enumerate}
 \item[Q1.] Are fragments with low complexity as useful as more complex fragments and if so, what are the underlying phenomena?
\end{enumerate} 
Restricting the number of fragments used gives us a controlled setting in which to evaluate the efficacy of fragments from different fragment classes. By analyzing the encoding of the data that can be derived from the mined patterns, and the relation among patterns themselves, we gain an intuition as to why simpler structures are the better choice when the number of patterns is limited in the mining process. 
It is not obvious whether these results will transfer to a size restricted setting but we can answer a related question, namely:
\begin{enumerate}
 \item[Q2.] Is mining fragments using a significance threshold as good as the size restricted approach to building \gfp{s}?
\end{enumerate}

The size restricted approach is equivalent to considering the occurrences of \emph{all} fragments adhering to those size restrictions and using those that occur at least once. This can lead to an explosion in the number of enumerated fragments, \emph{even} using length restriction on the patterns, as we will show. Arbitrarily increasing this threshold, on the other hand, might exclude interesting fragments, and as mentioned above, the effect of changing the size restrictions is not always predictable \cite{doi:10.1021/ci700052x}.

Analogously, minimally correlating fragments can be considered to consist of at least two atoms and to adhere to a correlation constraint. Changing the size constraint can still have unpredictable results whereas changing the correlation threshold has a clear interpretation. 
Consequently, in a final experiment, we compare the effect of increasing the number of fragments of low complexity by lowering the mining threshold, showing the increasing quality of the encoding (and its consequences for the quality of the classification model), and contrasting their usefulness with length restricted fragments.

\section{Approach\label{approach}}

We use substructures that correlate with one of two target classes (e.g. \emph{active} and \emph{inactive}) -- and therefore discriminate among the two. 
Techniques exist for mining
 top-$k$ substructures according to convex measures such as  $\chi^2$ or \emph{Information Gain} while still pruning large parts of the search space. Similar search strategies can be used to find all substructures with a score above a user defined threshold.
Please note that in this work we only use $\chi^2$ since earlier work showed that this leads to better results than employing Information Gain \cite{tree-squared}.

Regarding chemical compounds, there exist three very well studied types of substructures, namely:
\begin{enumerate}
\item[$\mathcal{L}_G$] \emph{subgraphs}, most expressive, but expensive to mine;
\item[$\mathcal{L}_T$] \emph{subtrees}, can represent anything but cycles;
\item[$\mathcal{L}_S$] \emph{subsequences}, least expressive, rather easy to mine.
\end{enumerate}

The relation $\mathcal{L}_S \subset \mathcal{L}_T \subset \mathcal{L}_G$ holds, implying that $|\mathcal{L}_S| \leq |\mathcal{L}_T| \leq |\mathcal{L}_G|$.
Note that sequences are slightly different from paths as used by \ea{Swamidass} as they only allow a bijective mapping of the nodes and edges from the fragment to the data, i.e. a vertex can occur at most once in a sequence.
Our first question is concerned with comparing these three types of structures w.r.t. their value in terms of predictive accuracy. To carry out this task, we  extract a number of substructures from the data, and use them to describe each of the seen or unseen chemical compounds. The molecules are 
 transformed into generalized fingerprints indicating the substructures' presence or absence.
From the feature vectors a model for the activity of the compounds is learned.

\emph{Support vector machines} (\algo{SVM}s) have been used successfully for SAR problems and can
filter out redundant/irrelevant features. We use the \emph{Tanimoto kernel} that has been used to good effect on the NCI cancer data set we do our comparison on \cite{baldi}. The data is encoded as undirected graphs, vertices labeled with their atom type, edges as single, double, or aromatic bonds. Hydrogen atoms are not encoded. The shortest possible sequence consists of a single edge, i.e. two atoms.

\section{Experimental Evaluation}
\label{experiments}

The NIC60 cancer data set is a popular data set for testing SAR predictions \cite{baldi,DBLP:conf/icml/CostaG10}. It consists of approximately 4000 compounds that have been tested against 60 tumor cell lines. Each of the 60 subsets consists of around 3,500 compounds.

For each of the 60 cell lines comprising the NCI cancer data set, we performed a stratified 10-fold cross validation.
 Fragments are mined exclusively on the training folds since there is no information about class labels in the test data.

\subsection{Comparing different complexity classes -- \fp{}s of equal size}

The classical approach to encoding molecules in \fp{}s lies in fixing the size $k$ of the \fp{} and hashing fragments to a bit-string of this length. In the technique we propose, mining significant fragments according to the $\chi^2$ statistic, fixing the size of \fp{}s can be done by mining only the $k$ highest-scoring fragments. In earlier work \cite{simpler-patterns}, we showed that for a fixed $k$, sequences were more effective in encoding data for classification purposes than trees or graphs. We repeat the experiments of this work here, this time using the NCI 60 data set used by \ea{Swamidass}.
For the mining process, $k$ was set to $1000$, the top-$k$ sequences, trees, and graphs mined and the AUC (area under the ROC) estimated as described above.

\begin{figure}
	\includegraphics[width=\linewidth]{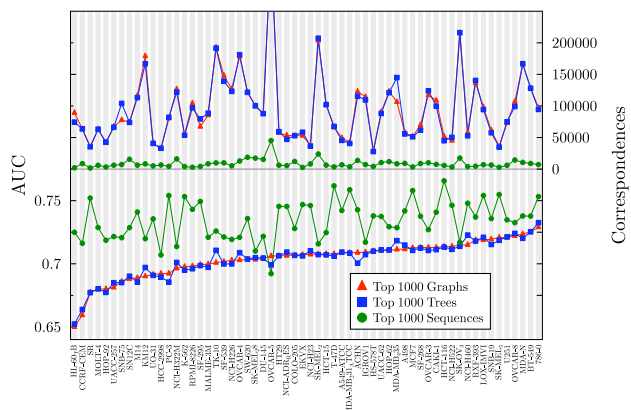}
	\caption{AUC results of SVM-classification on encodings derived from the top-1000 fragments and number of correspondences in each encoding.\label{top-1000-auc}}
\end{figure}

As the lower half of Figure \ref{top-1000-auc} shows, \fp{}s using sequences are significantly more useful for learning a model on the data using an SVM. This is consistent with our earlier results.
The \emph{Wilcoxon Matched Pairs Signed Rank Test} shows 39 out of 60 cases in which 
 using sequences is significantly ($p$-level $99\%$) better than
  using graphs and 38 out of 60 cases in which the tree-based encoding is outperformed.
To gain an understanding of the reasons for this, it is helpful to consider the encoding of molecules that can be derived from each mined set. 

We focus specifically on pairs of molecules from different classes that are encoded by the same \fp{} (correspondences). To give an idea of the prevalence of this phenomenon, we show for each fragment type and data set their average number 
 in the upper half of Figure \ref{top-1000-auc}.
In particular, this also includes all molecules that are not matched by any fragment at all. Such \fp{}s in the training data correspond to data points that the SVM cannot effectively learn to distinguish, while when unseen their classification is essentially up to chance. Generally speaking, the better performance of the SVM on sequence-encoded data aligns with fewer correspondences on this encoding.

\begin{figure}
	\centering
	\includegraphics[width=0.49\linewidth]{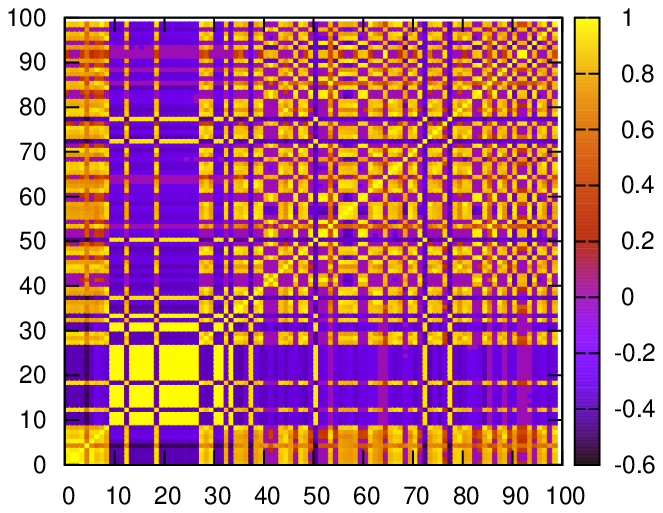}
	\includegraphics[width=0.49\linewidth]{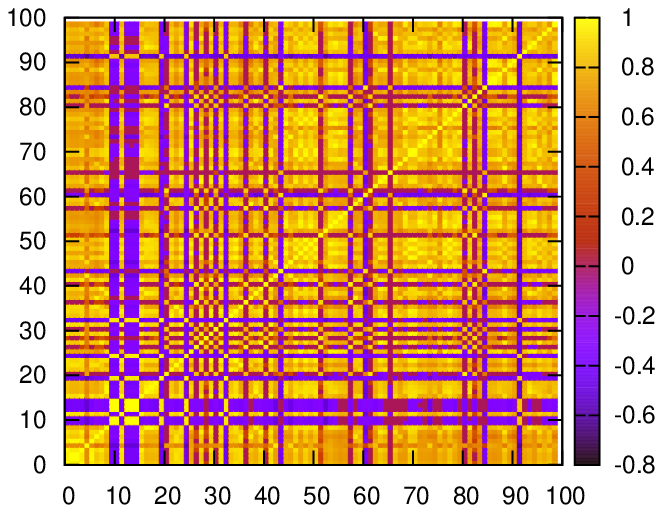}
	\caption[Intercorrelation of the $100$ best class-correlating graphs and sequences.]{\label{fig:pred_nci_corrtop100graph} Intercorrelation of the 100 best class-correlating graphs (sequences) in the left (right) correlation matrix. The sequences ranked 42 and lower are not in the top-$100$ graphs.}
\end{figure}

This phenomenon can be explained by the relationship of fragments to each other. 
Figure \ref{fig:pred_nci_corrtop100graph} shows the strength of correlation
of different fragments' presence.
As can be seen, sequences are much more diverse than graphs (with graphs and trees virtually identical), 
describing data more distinctively.

\begin{figure}
\begin{center}
	\includegraphics[width=0.9\linewidth]{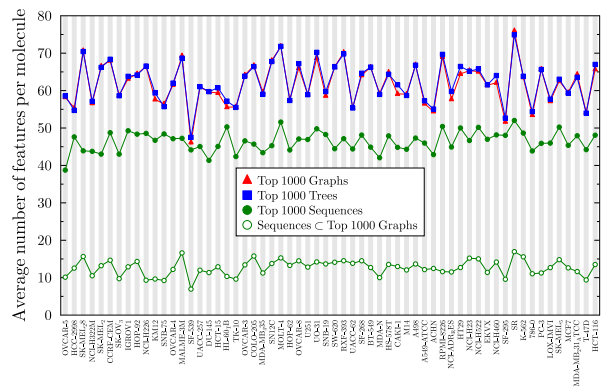}
	\end{center}
	\caption{Average number of \emph{features} per molecule for the top-1000 fragments.\label{top-1000-feat}}
\end{figure}

At first glance, the plot seems to disagree with this description since the left-hand side of the figure shows far less yellow, i.e. high pairwise correlations, than the right-hand side. However, 
in the figure, the lowest scoring sequence contained in the top-$100$ graphs corresponds to the sequence ranked $41$ in the top-$100$ sequences.
Hence, the lower-left part (up to fragment 41) of the top-$100$ sequences in the right is like a small scaled version of the left figure. This leads to the coarser appearance on the left-hand side, indicating that for each sequence there is at least one highly correlating graph. The most distinctive example for this can be found in the graphs ranked $\sim 8 - 26$: the majority of these graphs show very similar behavior w.r.t. pairwise correlation with other fragments and the entire block is equivalent to sequences ranked $\sim 8-15$. 

Note that these figures display only \emph{pairwise} correlation.
Using the final set of fragments during modeling allows for much more complex combinations than pairs.
Hence, these figures show only the tip of the iceberg, and we can expect the overall intercorrelation effects amongst combinations of fragments in the result set to be much stronger.

Figure \ref{top-1000-feat}, plotting the average number of features \emph{per molecule}, shows that the more complex fragments, while not reducing correspondences as efficiently,  use \emph{more} features. The lowest curve in this figure shows how few sequences are on average used for encoding once cyclic graphs crowd out less complex structures from the set.

\subsection{Comparing different complexity classes -- from \fp{}s to \gfp{}s}

\begin{figure}
\begin{center}
	\includegraphics[width=0.8\linewidth]{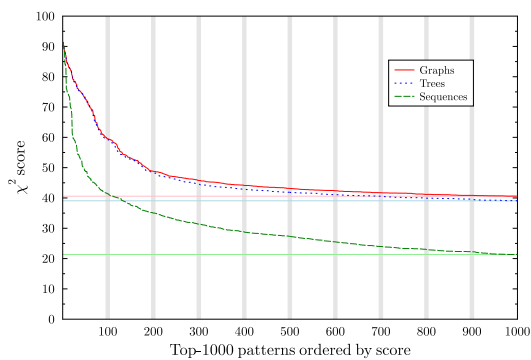}
\end{center}	
	\caption{Score distribution for the top-1000 fragments according to $\chi^2$ of one representative training fold.\label{top-1000-scores}}
\end{figure}

The higher diversity mentioned above also means that the scores of trees and graphs fall into a smaller interval than those of sequences.\footnote{In practice the best fragment is often a sequence. In general, the best graphs \emph{might} score much better than any sequences.} In other words, the $1000$th-best $\chi^2$ sequence score is lower than the $1000$th-best score for more complex structures (see Figure \ref{top-1000-scores}). The horizontal lines show the worst score for each structure type, which also indicates the number of fragments of other types that exceed it. 

\begin{figure}
\begin{center}
	\includegraphics[width=0.9\linewidth]{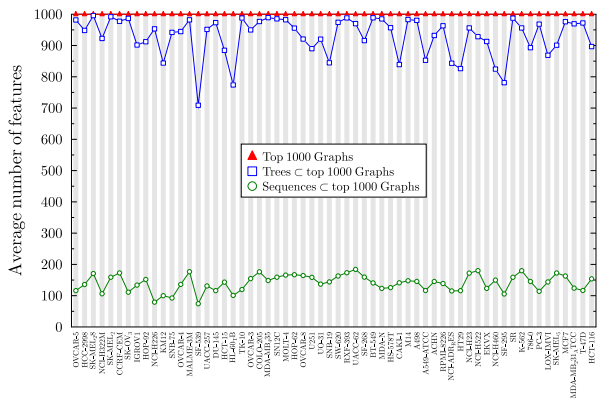}
\end{center}	
	\caption{Number of fragments remaining after the set is reduced using the 1000th-worst graph score.\label{cut-1000-afc}}
\end{figure}

To normalize the observed advantage of sequences with regard to diversity, we use the $1000$th-best graph score (the red horizontal line) to crop the size of \fp{}s derived from tree and sequential fragments. We effectively obtain \emph{generalized fingerprints}, without explicit size restriction of bit-vectors, similar to the ones used by \ea{Swamidas}, with the \emph{length} restriction on fragments replaced with a minimum \emph{significance} value. The number of fragments left is shown in Figure \ref{cut-1000-afc}.
This reduction is in fact rather severe, pushing the number of sequence fragments down to $10\% - 20\%$ of the original $1000$.

As mentioned before, Figure \ref{top-1000-feat} also shows the average number of sequential features per molecule in the top $1000$ graphs, which is equivalent to using the reduced set of sequential features (bottom curve).
In comparison to the second curve from the bottom, one can see that, again, the reduction is severe, yet not as severe as for the entire set of features, only dropping to $25\% - 33\%$ of the original number ($\sim 40-50$).

\begin{figure}
\begin{center}
	\includegraphics[width=0.9\linewidth]{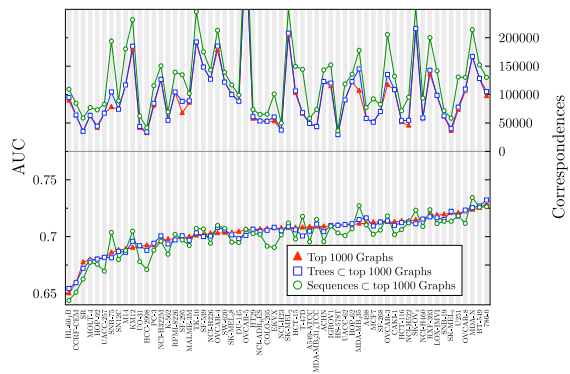}
\end{center}	
	\caption{AUC results of SVM-classification on encodings derived from the less complex fragments contained in the top-1000 graphs and number of correspondences in each encoding.\label{cut-1000-auc}}
\end{figure}

Reducing the number of features in this way leads to a very slight advantage for the graph-structured features w.r.t. AUC results
 (lower half of Figure \ref{cut-1000-auc}). In fact, according to the Wilcoxon test, there are only two significant differences,
 even though so many fewer 
 sequences than graphs are used.

Similarly to the results described in the preceding section, the decrease in accuracy goes along with an increase in the number of correspondences, as shown in the upper half of Figure \ref{cut-1000-auc}.
It is important to note the trade-off among the number of features and the quality of the feature set. The number of correspondences are rather similar for each data set for all structure types, despite the differences in total number of fragments. This indicates that the large sets of tree- and graph-structured fragments still show much redundancy, which in turn means that while additional complexity allows for some more diversity for a given threshold, the gain is relatively small compared to simply increasing the number of fragments.

\subsection{Increasing \gfp{}-size by lowering the mining threshold}

Increasing the number of features improves the chance that molecules from different classes are encoded in a way that allows to distinguish those classes. Tree and graph mining being far more expensive than sequcne mining \cite{simpler-patterns}, it is unrealistic to try and mine large amounts of complex fragments. Also, 
using all graphs which have a $\chi^2$-score exceeding a given threshold does little to increase diversity over sequences.

The computational complexity of fragment mining arises from the need to systematically explore a large search space of potentially interesting fragments and count their occurrences in the data. Approaches that start from individual molecules avoid this bottleneck, yet while the fragments derived in that manner can be used for assessing molecules' similarities this is often the extent of their usefulness, especially since they are often tied to their respective kernel functions. Fragments correlating with the target value, on the other hand, capture information about the relationship of structure and activity themselves and can be analyzed independently from the modeling step. 

\begin{figure}
\begin{center}
	\includegraphics[width=\linewidth]{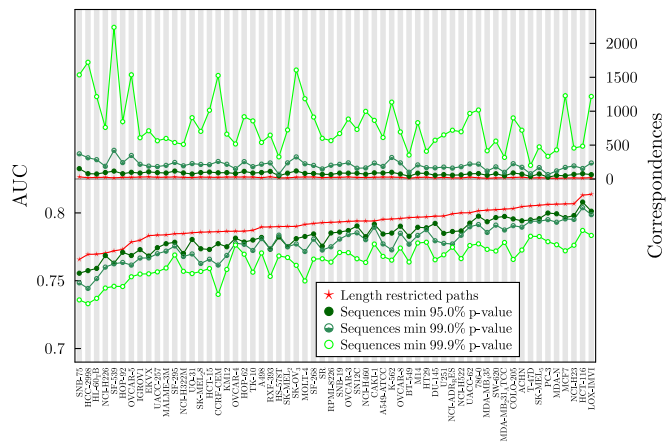}
	\end{center}
	\caption{AUC for sequences based on three different significance thresholds and length-restricted paths of frequency $1$ along with the number of correspondences in each encoding.\label{large-scale-aucs}}
\end{figure}

\begin{figure}
\begin{center}
	\includegraphics[width=0.95\linewidth]{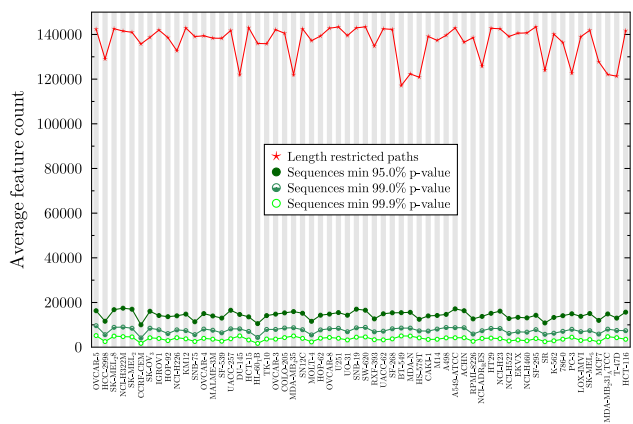}
\end{center}
	\caption{Number of sequential-fragments for three different significance thresholds and length-restricted paths of frequency $1$.\label{large-feature-counts}}
\end{figure}

In a third experiment 
we thus mine sequences which have a $\chi^2$-score exceeding the (unadjusted) $95\%$, $99\%$, and $99.9\%$ p-values, respectively. As expected, Figure \ref{large-feature-counts} shows  a direct relationship between lowering the significance threshold and the number of features mined. 
More features also leads to fewer correspondences which lead then to corresponding AUC values (Figure \ref{large-scale-aucs}).
According to the Wilcoxon Test, using the p-value for $99\%$ improves significantly on the $99.9\%$ value in 36 cases. Using the $95\%$ value increases this to 45 (and improves in 17 cases on the $99\%$ value).

\subsection{Comparing techniques for determining \gfp{}-size: significance \emph{versus} length-restriction}

The preceding experiments had the main purpose of assessing the usefulness of graph-, tree- and sequential fragments for \gfp{}s, chosen by the significance of their $\chi^2$ score. \ea{Swamidass} \cite{baldi} use \gfp{}s whose number is determined by a length restriction -- all fragments are paths of maximal length $10$ occuring in at least \emph{one} molecule in the training data. This approach gives rise to more than one hundred  thousand features (the top graph in Figure \ref{large-feature-counts}), significantly more than result even from using the permissive $95\%$ p-value. According to the results described above, this should allow those feature sets to encode all molecules distinctively. Indeed, the low frequency threshold means that there are very few correspondences, as can be seen the top part of Figure \ref{large-scale-aucs}. While the number of correspondences is reduced even further, however, they are not eliminated completely -- it would probably need longer paths (and thus many more fragments) to effect this. 



In Swamidass's work two kernels are used for classification -- one based on the well-established Tanimoto-similarity \cite{tanimoto} and a so-called \emph{Min-Max-Kernel}. While the latter gives slightly better results w.r.t. predictive accuracy, it is evaluated on a representation of the data that not only denotes absence/presence of substructures but also the number of times they occur in a molecule. Since the semantic information of significant fragments is such that only their \emph{presence} correlates with an activity, we do not adopt this representation and thus do not compare against the Min-Max-Kernel.

\begin{figure}
\begin{center}
	\includegraphics[width=0.9\linewidth]{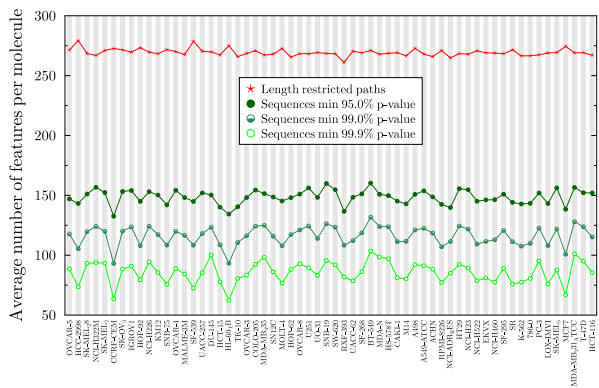}
\end{center}
	\caption{Average number of fragments per molecule mined with three different significance thresholds and as length restricted paths of frequency $1$.\label{large-scale-avg-count}}
\end{figure}

The lower half of Figure \ref{large-scale-aucs} also lists the average AUC the SVM achieved on \gfp{}s using length-restriction. As we expected, using length-restricted paths with minimum support of one leads to slightly more useful feature sets but at the cost of significantly larger \fp{}s. In fact, while the AUC increase is not significant for most data sets (only 13/60 according to the Wilcoxon test), Figure \ref{large-scale-avg-count} shows that the average number of fragments used to describe a single molecule effectively doubles. The gain derived from increasing the amount of features is thus affected by diminishing returns.

\begin{figure}
\begin{center}
	\includegraphics[width=0.9\linewidth]{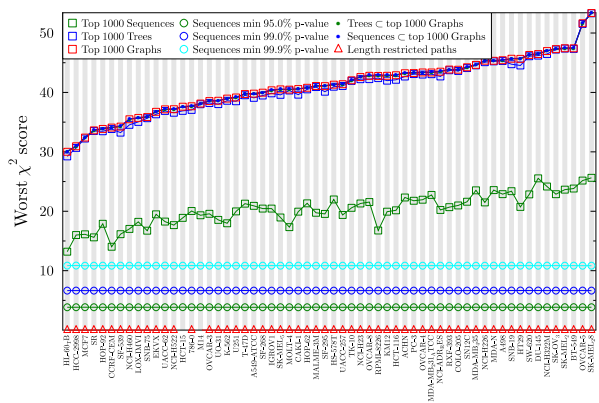}
\end{center}
	\caption{Worst score for fragments mined with three different significance thresholds and as length restricted paths of frequency $1$.\label{large-scale-min-scores}}
\end{figure}

Those fragments are relevant in terms of the similarity of molecules yet do not capture any tendencies in the data themselves. The \emph{kernel matrix} gives a global view of the similarity of molecules and the SVM is used to discover the underlying phenomena. Figure \ref{large-scale-min-scores} shows that the lowest scores for fragments derived in the length-restricted approach are clearly non-significant and it would be hard to base actual understanding of the data on them. It also shows that the scores of the worst sequence and worst tree included in the top-thousand graphs are virtually indistinguishable from each other and from what is considered the worst graph. Finally, the score of the 1000th-worst tree is marginally worse than the score of the 1000th-worst graph, indicating that most of the top-1000 graphs are in fact trees.

\section{Conclusions}
\label{conclusion}

We performed an empirical evaluation to gain insight into the reasons for the superiority of sequential molecular fragments as features for classification,  compared 
to complex ones. 
We find  that the reason lies in the greater diversity of sequences which leads to a more distinctive encoding of instances, effectively giving classifiers a better representation to work with. A straight-forward way of improving the encoding lies in increasing the number of fragments used.
Our experiments show, however, that there is \emph{always} need for a far greater number of trees/graphs than sequences. As these structures are also computationally more expensive  to enumerate, 
this leads to vastly increased computational complexity. Our results indicate that this should be avoided.

Enumerating a subset of all sequences that cover at least one molecule produces an effective feature set but also a very large one. Also, these fragments are hard to interpret outside of their use in a pairwise similarity measure. In contrast to this, fragments that are selected based on their correlation with the target can be ranked based on their score and the most interesting ones inspected and interpreted by an end user. While it would be possible to evaluate all size restricted fragments on the data and perform a similar ranking, this will be computationally expensive due to their large number.


Our experiments also indicate a clear trade-off between the number of fragments (which can be set by adjusting the $k$ in top-$k$ mining or the minimum significance threshold) and the quality of the feature set. Given existing results, it is unlikely that similar clear-cut effects would appear when changing the  length or minimum support of length restricted paths.

Redundancy among complex patterns could be reduced explicitly, e.g. in a post-processing step. We have suggested a technique that achieves this \cite{chosen-few} and since the fragments can be considered features, 
 \emph{feature selection} techniques are applicable \cite{DBLP:journals/jmlr/GuyonE03}, but this would again require the mining of a \emph{large} set of trees or graphs. It would need to be larger than a set of sequential patterns that could be used without post-processing, 
  increasing computational complexity significantly. 

Given these arguments, class-correlated sequences seem to be the best choice for the large-scale mining of molecular fragments as features for SAR prediction.

An alternative lies in iterative approaches, in which patterns are mined and data manipulated \cite{tree-squared,DBLP:conf/sdm/Thoma09,DBLP:conf/pkdd/ZimmermannBR10}, or redundancy with already found patterns made part of the quality function \cite{ruckertK07}. Effective, very compact pattern sets can be mined in this way. 
So far there is however no clear understanding about whether these feature sets would be competitive with large sets of sequences for classification. Additionally, due to sequences being graphs themselves, as explained in the introduction, it seems quite possible that such a mining operation would in the end once again give rise to a set of sequential fragments.

\section*{Acknowledgements}
We would like to thank Kurt De Grave for his invaluable help in preparing this manuscript. Albrecht Zimmermann was supported by the Fonds Wetenschappelijk Onderzoek (FWO) by the time of writing.



\bibliographystyle{ieeetr}
\bibliography{../bibliographie}
\end{document}